\PassOptionsToPackage{table}{xcolor}

\documentclass[nonacm, sigconf, balance = false]{acmart}

\usepackage{booktabs} 

\acmJournal{TOG}

\copyrightyear{2023} 
\acmYear{2023} 
\setcopyright{acmlicensed}
\acmConference[SIGGRAPH Asia '23 Conference Proceedings]{Special Interest Group on Computer Graphics and Interactive Techniques Asia Conference Conference Proceedings}{December 12--15, 2023}{Sydney, Australia}
\acmBooktitle{Special Interest Group on Computer Graphics and Interactive Techniques Asia Conference Conference Proceedings (SIGGRAPH Asia '23 Conference Proceedings), December 12--15, 2023, Sydney, Australia}
\acmPrice{15.00}
\acmDOI{XX.XXXX/XXXXXXXX.XXXXXXXX}
\acmISBN{XXX-X-XXXX-XXXX-X/XX/XX}



\setcitestyle{square}

\usepackage[mathletters]{ucs}

\newcommand{\refequ}[1]{(\ref{equ:#1})}

\newcommand{\reffig}[1]{Fig.~\ref{fig:#1}}

\definecolor{LavenderBlue}{rgb}{0.69,    0.87,    0.54}
\usepackage{array}
\providecommand{\A}{}
\providecommand{\B}{}

\providecommand{\M}{}

\providecommand{\P}{}
\providecommand{\Q}{}
\providecommand{\R}{}
\providecommand{\S}{}

\providecommand{\V}{}

\providecommand{\a}{}

\providecommand{\c}{}

\providecommand{\p}{}

\providecommand{\t}{}

\usepackage{tabularx}
\usepackage{booktabs}
\usepackage{makecell}
\usepackage[table]{xcolor}
\definecolor{white}{rgb}{1,1,1}
\definecolor{lightbluishgrey}{rgb}{0.76471,0.84824,0.91647}

\usepackage{subcaption}
\usepackage{wrapfig}
\usepackage{graphicx}

\usepackage{listings}
\usepackage{layouts}
\makeatletter 
\newcommand{\layoutdetails}{\begin{tabular}{ll}
 \texttt{\textbackslash{textwidth}} & \printinunitsof{in}\prntlen{\textwidth} \\
\texttt{\textbackslash{linewidth}} & \printinunitsof{in}\prntlen{\linewidth} \\
Main text font &  \f@size pt \f@family \\
\sffamily \small Caption text font &  \sffamily \small \f@size pt \f@family \\
\end{tabular}}
\makeatother

\DeclareMathOperator{\clamp}{clamp}

\DeclareMathOperator*{\argmin}{argmin}
\DeclareMathOperator{\sd}{\phi}

\DeclareMathOperator{\ESDF}{\mathcal{E}_{\sd}}
\renewcommand{\p}{\mathbf{p}}
\renewcommand{\P}{\mathbf{P}}

\renewcommand{\c}{\mathbf{c}}

\renewcommand{\M}{\mathbf{M}}
\renewcommand{\A}{\mathbf{A}}
\renewcommand{\a}{\mathbf{a}}

\renewcommand{\t}{\mathbf{t}}
\renewcommand{\S}{\mathbf{S}}
\renewcommand{\Q}{\mathbf{Q}}
\renewcommand{\B}{\mathbf{B}}
\renewcommand{\R}{\mathbb{R}}

\begin{document}
\title[\emph{Reach For the Spheres}: Tangency-Aware Surface Reconstruction of SDFs]{\emph{Reach For the Spheres}: \linebreak Tangency-Aware Surface Reconstruction of SDFs}

\author{Silvia Sell\'{a}n}
\affiliation{\institution{University of Toronto}\country{Canada}}
\email{sgsellan@cs.toronto.edu}

\author{Christopher Batty}
\affiliation{\institution{University of Waterloo}\country{Canada}}
\email{christopher.batty@uwaterloo.ca}

\author{Oded Stein}
\affiliation{\institution{University of Southern California}\country{United States of America}}
\email{ostein@usc.edu}

\begin{abstract}
Signed distance fields (SDFs) are a widely used implicit surface representation, with broad applications in computer graphics, computer vision, and applied mathematics. To reconstruct an explicit triangle mesh surface corresponding to an SDF, traditional isosurfacing methods, such as Marching Cubes and and its variants, are typically used. However, these methods overlook fundamental properties of SDFs, resulting in reconstructions that exhibit severe oversmoothing and feature loss. To address this shortcoming, we propose a novel method based on a key insight: each SDF sample corresponds to a spherical region that must lie fully inside or outside the surface, depending on its sign, and that must be tangent to the surface at some point. Leveraging this understanding, we formulate an energy that gauges the degree of violation of tangency constraints by a proposed surface. We then employ a gradient flow that minimizes our energy, starting from an initial triangle mesh that encapsulates the surface. This algorithm yields superior reconstructions to previous methods, even with sparsely sampled SDFs. Our approach provides a more nuanced understanding of SDFs and offers significant improvements in surface reconstruction.
\vspace{60pt}
\end{abstract} 
\begin{teaserfigure}
\vspace{15pt}
    \centering
\includegraphics[width=\textwidth]{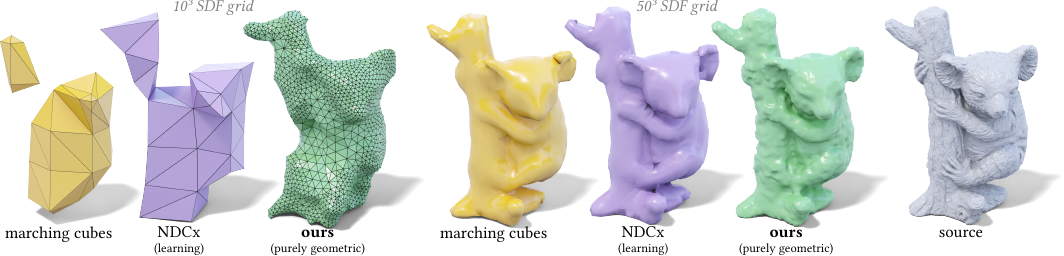}
\caption{
    Reconstructing a mesh from the discrete signed distance field (SDF) of a koala (source, \emph{rightmost}).
    By using global information from all sample points at once, our method recovers the shape even in low resolutions where methods like Marching Cubes
    \cite{lorensen1987marching}
    and Neural Dual Contouring (NDCx)
    \cite{chen2022neural}
    produce very coarse shapes (\emph{left trio}), and it recovers surface detail at higher resolutions that Marching Cubes and NDCx miss (\emph{middle trio}).
    Our method is purely geometric, and does not require any training or storing of weights (unlike NDCx).
    \vspace{30pt}
}\label{fig:teaser}
\end{teaserfigure}

\maketitle

\begin{figure}
\includegraphics{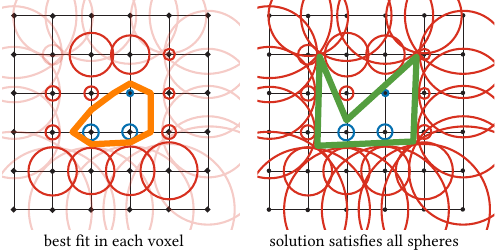}
\caption{Reconstructing an SDF per-voxel, by finding a best fit line segment in each voxel containing both positive (red) and negative (blue) values, discards much of the available global information.
    Our main insight is that a solution adhering to the constraints of \emph{all} spheres yields better results.}\label{fig:insight}
\end{figure}

\begin{figure}
\includegraphics{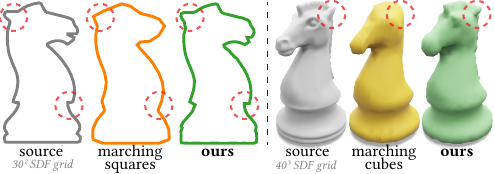}
\caption{Using global information (not just per-voxel data), we reconstruct sharp features even at low resolutions.
}\label{fig:sharp-features}
\end{figure}

\section{Introduction}
Signed distance fields (SDFs) are a classical implicit surface representation that finds diverse applications in computer graphics, computer vision, and applied mathematics, among other domains \cite{frisken2000adaptively,sethian1999level,jones20063d}.
A continuous SDF is a scalar function $\sd(\mathbf{x})$ that, given a query point $\mathbf{x}$ in $\mathbb{R}^n$, returns the Euclidean distance to the closest point on the surface it represents, augmented with a sign indicating whether the point is on the interior or exterior. A discrete SDF samples this function at a finite set of points in space, such as a grid, octree, or point cloud. The task we consider is the reconstruction of an explicit triangle mesh corresponding to the zero isosurface of such a discrete SDF.

Perhaps the most familiar such isosurfacing approach is Marching Cubes \cite{lorensen1987marching} and its variants. They use sign changes between adjacent SDF samples (e.g., along grid edges) to approximately locate the zero isosurface and apply per-cell templates and linear interpolation of the function values to fill in local patches of the surface triangulation. While effective and appropriate for \emph{general} implicit surface data, these schemes ignore the unique and fundamental properties of SDFs to their detriment. Indeed, mesh reconstructions of SDF data invariably exhibit severe oversmoothing and feature loss (see \reffig{teaser}). Approaches like dual contouring \cite{kobbelt2001feature,ju2002dual} can better recover sharp features by additionally relying on surface normals. However, discrete SDFs lack the necessary gradient information and finite difference estimates give disappointing results. Is there any hope of achieving better reconstructions from the SDF data alone?

Neural Marching Cubes \cite{chen2021neural} and Neural Dual Contouring  \cite{chen2022neural} have recently answered this question in the affirmative: they demonstrate better per-cell reconstructions by training on a large dataset of SDFs and using wider ($7^3$) stencils of SDF grid points. The quality of these results suggests that there exists some additional information implicit in the SDF data. Our objective is therefore to explicitly identify and directly exploit this overlooked geometric information, without recourse to learning approaches, and thereby achieve superior reconstructions.

The key insight underpinning our method is that (see Fig. \ref{fig:insight}) \emph{each SDF sample $\sd(\mathbf{x})$ corresponds to a spherical region}, centered at $\mathbf{x}$ and with radius equal to $|\sd(\mathbf{x})|$. By definition, the true surface represented by the discrete SDF must be tangent to every sphere at least once while strictly containing every sphere with negative value and excluding every positive value one. Through these constraints, the SDF samples contain significantly more information about the surface than samples from a generic implicit representation would.

To fully exploit this information, we first formulate an energy that measures the degree to which a proposed surface violates the tangency constraints of the input SDF samples. We propose an algorithm that starts from a triangle mesh enclosing the surface, then "shrinkwraps" the underlying surface via a gradient flow that minimizes our energy, interleaved with remeshing to ensure mesh quality. 
The fidelity of the resulting reconstructions surpasses that of prior methods, especially for sparsely sampled SDFs. Additionally, since our method has no intrinsic dependence on a grid, it is amenable to unstructured point cloud SDFs, and even incorporating new samples, where available, to improve the reconstruction.

\begin{figure}
\includegraphics{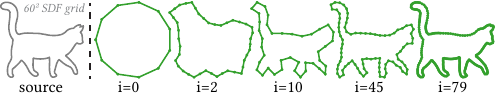}
\caption{Our 2D flow at different iteration counts on a cat,
    sampled from the source mesh on the \emph{left}.
}\label{fig:flow-2d}
\end{figure}

\begin{figure}
\includegraphics{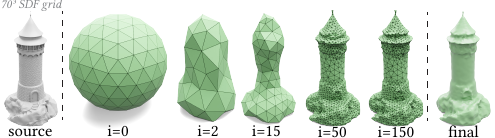}
\caption{Our 3D flow at different iteration counts on a tower,
    sampled from the source mesh (\emph{left}).
    The final version (\emph{right}) has been Loop subdivided.
}\label{fig:flow-3d}
\end{figure}

\section{Related Work}

\subsection{Signed Distance Fields}
SDFs have been used in countless applications spanning the computational sciences, so we highlight only a representative sample.
In computer graphics they have been applied to liquid surface tracking \cite{foster2001practical}, geometric modeling \cite{museth2002level}, collision detection \cite{fuhrmann2003distance}, and ray (sphere) tracing \cite{hart1996sphere}.
In traditional computer vision, uses have included image / volume segmentation \cite{chan1999active} and surface reconstruction from multiview data \cite{faugeras1998} or point clouds \cite{zhao2001fast}.
In computational physics, SDFs have been applied (via the level set method) to model combustion, crystal growth, and fluid dynamics \cite{sethian1999level,osher2004level}.
SDFs have also been applied to manufacturing \cite{Brunton2021} and robot path planning \cite{Liu2022}.

\begin{figure*}
\includegraphics[width=\linewidth]{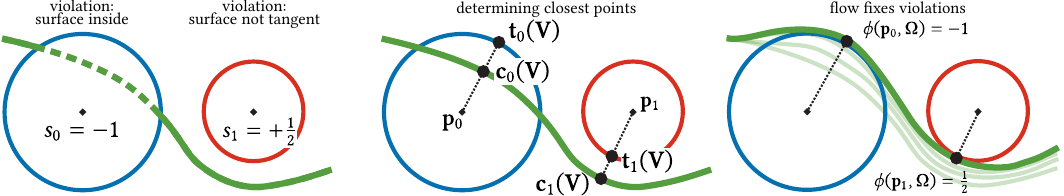}
\caption{A green surface the sphere constraints from two SDF samples \emph{(left)}.
    The method identifies the closest point on the surface and the closest point on the sphere for each sample point \(\p_i\) that makes the sphere correctly lie inside or outside the surface, as prescribed by \(s_i\) \emph{(middle)}.
    Our flow fixes all violations until the constraints \(\sd(\p_i,\mathcal{M}) = s_i\) are fulfilled \emph{(right)}.}\label{fig:didactic-flow-2d}
\end{figure*}

\begin{figure}
\includegraphics{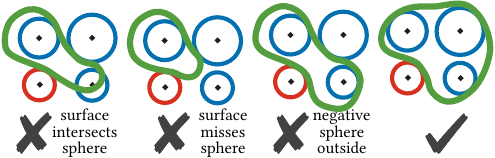}
\caption{Three surfaces violating the SDF constraints in different ways, and a surface that satisfies them \emph{(left to right)}.}\label{fig:didactic-violations}
\end{figure}

SDFs have recently seen renewed interest in the context of geometric deep learning. In particular, the DeepSDF approach \cite{park2019deepsdf} replaces the discrete SDF with a learned continuous SDF of a shape or a space of shapes.
This concept represents a subset of general neural implicit surfaces and of neural fields even more broadly \cite{neuralfields}.
Differentiable rendering with SDFs has been investigated to solve inverse problems \cite{bangaru2022differentiable,vicini2022differentiable}.
Since neural implicits often fail to exactly retain the signed distance property, \citet{sharp2022spelunking} propose techniques for robust geometric queries in this setting.

Despite
widespread adoption of SDFs, prior techniques often view SDFs as a convenient and canonical but \emph{general} implicit surface function.
They may exploit the distance property in some respects, but none that we are aware of consider the additional subgrid information implied by our tangent-spheres interpretation (mentioned by \citet{kobbelt2001feature,battyblog}).
Instead, the location of the
zero
isosurface is
assumed to be that of the linear (or occasionally polynomial) interpolant of the SDF samples.
However, away from the data points, such interpolated fields are not true SDFs.

\subsection{Isosurfacing Approaches}
The task of generating an explicit mesh corresponding to a given implicit surface is variously referred to as isosurfacing, polygonization, surface reconstruction, or simply meshing. There is an extensive body of literature on the topic, so we refer the reader to the survey by \citet{de2015survey}. There exist three broad categories of isosurfacing schemes: first, spatial subdivision schemes like Marching Cubes; second, advancing front methods, which start at a point and incrementally attach new triangles as they propagate across the surface until it is covered
\cite{hilton1996marching,Sharf2006}; and third, shrinkwrap or inflation methods, which start with an initial closed surface and gradually grow it inwards or outwards to conform to the desired isosurface \cite{van2004shrinkwrap,stander1997,Hanocka2020}. Our method falls into the last category, which is relatively under-explored compared to the others.
The work of \citet{Bukenberger2021} employs evolving meshes with periodic remeshing that conform to target SDFs, like our approach.
They require the SDF to be resampled at every iteration of their method, while our method only requires the SDF to be evaluated on a finite set of evaluation points at the beginning (although our method can support resampling as well, see Fig.~\ref{fig:resampling}), and they do not use all SDF spheres' global information.

Isosurfacing methods are often applied to SDFs, but seldom exploit the signed distance property. Indirectly, Neural Marching Cubes \cite{chen2021neural} and Neural Dual Contouring \cite{chen2022neural} represent two key exceptions. Their SDF dependence is not explicit in their algorithms, but implicitly encoded into their neural networks when trained on exact SDFs to achieve improved results compared to prior non-neural schemes. Our approach avoids any reliance on deep learning, instead making explicit use of fundamental geometric properties of SDFs. 

Recently, Deep Marching Cubes (DMC) \cite{liao2018deep}, MeshSDF \cite{remelli2020meshsdf},  and \textsc{FlexiCubes} \cite{Shen2023}  were developed to offer \emph{differentiable} isosurfacing procedures. Their goal is often to incorporate isosurfacing into end-to-end deep learning pipelines for applications like shape completion, shape optimization, or single-view reconstruction. By contrast, we focus on achieving the highest quality of reconstruction of discrete SDFs.  Our SDF energy has connections to the losses used in such work.

\subsection{Mesh Optimization}
Our algorithm uses gradient flow to optimize an energy with respect to the vertex positions of a triangle mesh, with the aim of finding a valid, tangency-aware surface reconstruction.
Variational techniques in this style are ubiquitous in geometry processing applications, such as mean curvature flow and surface fairing \cite{desbrun1999implicit,Kazhdan2012}, mesh quality improvement \cite{alliez2005variational}, Willmore flow \cite{crane2013robust}, constructing coarse cages \cite{Sacht2015}, developability \cite{stein2018developability}, and morphological operations \cite{sellan2020opening}.
To maintain and improve mesh quality during our flow, we employ the local remeshing scheme of \citet{botsch2004remeshing}.

Our flow displaces the vertices of a mesh such that the mesh is tangent to a set of spheres centered at the SDF sample points. In a way, this can be seen as solving a reverse formulation of the medial axis computation problem. In it, one searches for maximally contained spheres tangent to a given surface, often through a combination of greedy decompositions and progressive simplifications \cite{li2015q,ma20123d,rebain2019lsmat}.

\begin{figure}
\includegraphics{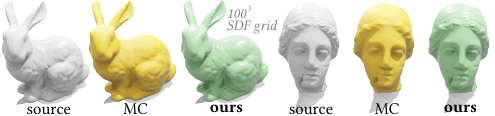}
\caption{While our algorithm shines at low and medium resolutions, it also recovers high-frequency detail Marching Cubes misses at higher resolutions.
}\label{fig:high-resolution}
\end{figure}

\section{Method}

Let us assume we are given access to the values $s_1,\dots,s_n\in\R$ of the Signed Distance Function $\sd$ for an unknown surface $\Omega$ sampled at $n$ points in space $\p_1,\dots,\p_n\in\R^3$, $s_i = \sd(\p_i, \Omega)$. Our task will be to reconstruct a \emph{valid} surface $\Omega$; i.e., one that is consistent with the SDF observations.
This can be expressed as the constraints
\begin{equation}\label{equ:sdfsmoothconstraint}
    \sd(\p_i,\Omega) = s_i \,, \quad \forall i \in \{1,\dots,n\} \,.
\end{equation}
Intuitively, one may visualize this condition by drawing a sphere $S_i$ of radius $|s_i|$ around each $\p_i$ and requiring that the surface $\Omega$ be tangent to all of them with the correct orientation (see Fig.~\ref{fig:didactic-violations}).

We begin with a simple idea: turn \refequ{sdfsmoothconstraint} into an energy minimization problem over the space of surfaces. To this end, we define the \emph{SDF energy} of a surface to be the squared difference between $s_i$ and the SDF value of the surface at $\p_i$:
\begin{equation}\label{equ:sdferror}
    \ESDF(\Omega) = \frac{1}{2}\sum_{i=1}^n \left(\sd(\p_i,\Omega) - s_i\right)^2 \,.
\end{equation}
Exploring the entire space of surfaces $\Omega$ to find one that minimizes the above energy is intractable. Instead, we propose to start from an initial surface $\Omega^0$ and follow the gradient flow of the SDF energy
\begin{equation}
    \frac{\partial \Omega}{\partial t} = -\nabla \ESDF(\Omega)
\,.
\end{equation}

We refer to this as our \emph{sphere reaching} flow, since it will encourage $\Omega^t$ to touch every $S_i$ at least once, while strictly containing all negative $S_i$ and strictly excluding all positive $S_i$ (see Fig.~\ref{fig:didactic-flow-2d}).

\section{Discretization}

We discretize the time dimension of our flow using an implicit scheme to obtain the sequence $\Omega^0,\Omega^1,\dots,\Omega^T$ of surfaces such that
\begin{equation}
    \Omega^t =  \Omega^{t-1}-\tau\nabla \ESDF\left(\Omega^t\right) \,
\end{equation}
for some small time step $\tau$. Equivalently, given a surface $\Omega^{t-1}$, we will aim to find a new surface $\Omega^t$ that minimizes the energy
\begin{equation}\label{equ:timediscretized}
    \Omega^{t} = \argmin_{\Omega} \frac{1}{2\tau} \left\|\Omega - \Omega^{t-1}\right\|_2^2 + \ESDF(\Omega)  \,.
\end{equation}

In order to sequentially solve this minimization problem for $t=0,\dots,T$, we will need to discretize the space of surfaces $\Omega$. We will do so by representing each surface $\Omega^t$ as a triangle mesh \(\mathbf{\Omega}^t\) with vertices $\mathbf{V}^t$ and faces $\mathbf{F}^t$. \refequ{timediscretized} can then be written as
\begin{equation}\label{equ:timediscretized2}
    \Omega^{t} = \argmin_{\Omega} \left( \frac{1}{2\tau} \left\|\mathbf{V} - \mathbf{V}^{t-1}\right\|_\M^2 + \ESDF(\Omega) \right) \,,
\end{equation}
where $\|\cdot\|_\M$ is the mass-matrix-integrated norm $\|\cdot\|^2_\M = \cdot^\top\M\cdot$.

Unfortunately, $\ESDF(\mathbf{\Omega})$ is not convex or even continuously differentiable, which makes it difficult to minimize. 
To circumvent this, we first define $\c_i(\mathbf{\Omega})$ as the specific closest point on the surface $\mathbf{\Omega}$ to $\p_i$,\footnote{We use an AABB tree to speed up computation of the closest point, and we compute the query for every \(p_i\) at once.} which we can write as $\a_i({\mathbf{\Omega}}) \mathbf{V}$ for some sparse vector of barycentric coordinates $\a_i({\mathbf{\Omega}})$.
Then $\ESDF(\mathbf{\Omega})$ becomes
\begin{equation}\label{equ:sdfenergydiscrete}
    \ESDF(\mathbf{\Omega}) = \frac{1}{2}\sum_{i=1}^n \left(\sd(\p_i,\c_i(\mathbf{\Omega})) - s_i\right)^2
\end{equation}
where we have slightly abused notation to let $\sd(\p_i,\c_i(\Omega))$ return the distance $\|\p_i - \c_i(\Omega)\|$ with the sign of $\sd(\p_i,\mathbf{\Omega})$. This modified energy penalizes distances between each closest point and the corresponding sphere's surface. Refer to  Fig.~\ref{fig:didactic-flow-2d} for the geometric picture.

\begin{figure}
\includegraphics{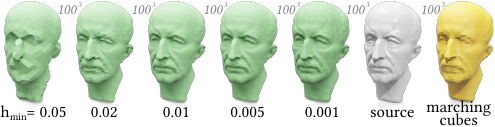}
\caption{A critical parameter in our method is the maximum resolution, encoded in the  edge-length $h_\textrm{min}$, which balances the under- or over-constrained nature of our optimization.}\label{fig:hyperparameters}
\end{figure}

\begin{figure*}
\includegraphics[width=\linewidth]{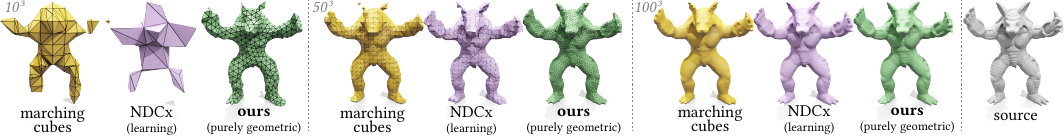}
\caption{
    With no training and no network weight storage, our method consistently outperforms MC and outperforms or matches Neural DC across resolutions.
}\label{fig:armadillo}
\end{figure*}

We next define \(\t_i(\mathbf{\Omega})\) as the projection of $\p_i$, along the line through $\p_i(\mathbf{\Omega})$ and $\c_i(\mathbf{\Omega})$, onto the signed distance sphere $S_i$,
\begin{equation}\label{equ:ti}
    \t_i(\mathbf{\Omega}) =
    \p_i + \sigma_i |s_i|
    \frac{\c_i(\mathbf{\Omega})-\p_i}{\left\lVert
    \c_i(\mathbf{\Omega})-\p_i
    \right\rVert}
    \;,
\end{equation}
where \(\sigma_i\) depends on the orientation of the surface \(\mathbf{\Omega}\) at \(\c_i(\mathbf{\Omega})\):
\begin{itemize}
    \item If \(\mathbf{p}_i\) is inside/outside \(\mathbf{\Omega}\) and the sign of \(s_i\) is negative/positive, then \(\sigma_i=1\).
    \item If \(\mathbf{p}_i\) is inside/outside \(\mathbf{\Omega}\) and the sign of \(s_i\) is positive/negative, then \(\sigma_i=-1\).
\end{itemize}
That way, if the surface were translated such that \(\c_i(\mathbf{\Omega})\)
coincided with \(\t_i\), the SDF would be satisfied at \(\p_i\).
We use the mesh element's normal vector at \(\c_i(\mathbf{\Omega})\) to distinguish between inside and outside.\footnote{In practice, we do not distinguish between inside and outside spheres if \(\c_i(\mathbf{\Omega})\) is on the boundary of a mesh element, since normal information is not reliable there --- \(\t_i(\Omega)\) is simply the closest point on the sphere.}

\begin{figure}
\includegraphics{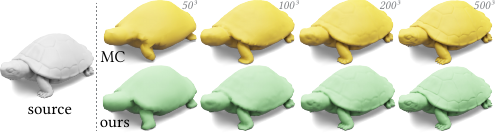}
\caption{
    Our flow can be used on sparse (recovering more detail than MC) and very dense (matching MC) SDF grids.}\label{fig:large-grid}
\end{figure}

As $\t_i(\mathbf{\Omega})$ and $\c_i(\mathbf{\Omega})$ will be equal for a valid solution, we approximate \(\left( \sd(\p_i,\c_i(\mathbf{\Omega})) - s_i \right)\)
by \(\left\lVert \c_i(\mathbf{\Omega}) - \t_i(\mathbf{\Omega})) \right\rVert\). Thus, (\ref{equ:sdfenergydiscrete}) becomes
\begin{equation}\label{equ:sdfenergydiscrete2}
    \frac{1}{2}\sum_{i=1}^n \left\lVert \c_i(\mathbf{\Omega}) - \t_i(\mathbf{\Omega})) \right\rVert^2
    =
    \frac{1}{2}\sum_{i=1}^n \left\lVert \a_i({\mathbf{\Omega}}) \mathbf{V} - \t_i(\mathbf{\Omega}) \right\rVert^2
    \,,
\end{equation}
which we further simplify by fixing \(\t_i(\mathbf{\Omega})\)
to \(\t_i\left(\mathbf{\Omega^{t-1}}\right)\).

Concatenating $\a_i$ and $\t_i$ into matrices $\A$ and $\S$, this becomes
\begin{equation}
    \frac{1}{2}\|\A \mathbf{V} - \S\|_F^2\,,
\end{equation}
which we can now incorporate into \refequ{timediscretized2}:
\begin{equation}
    \mathbf{V}^{t} = \argmin_{\mathbf{V}} \frac{1}{2\tau} \left\|\mathbf{V} - \mathbf{V}^{t-1}\right\|_\M^2 + \frac{1}{2}\|\A \mathbf{V} - \S\|_F^2\,.
\end{equation}
This quadratic optimization problem on the vertex positions $\mathbf{V}$ can be solved via the linear system
\begin{equation}\label{eq:discretetimestepping}
    \Q \mathbf{V}^t = \B
\end{equation}
where
\begin{equation}
    \Q = \M + \tau \A^\top\A\,,\qquad \B = \M\mathbf{V}^{t-1} + \tau\A^\top\S\, .
\end{equation}
The matrices \(\mathbf{A}\), \(\mathbf{M}\), and \(\mathbf{Q}\) are sparse, thus the linear system can be efficiently solved using, e.g., Cholesky decomposition.
A simplified step of this flow can be seen on the right of
Fig.~\ref{fig:didactic-flow-2d}.

\paragraph{Choosing step size \(\tau\).}
The formulation in (\ref{eq:discretetimestepping}) produces a flow that, for a small enough $\tau$, will reduce the SDF energy each iteration. However, choosing $\tau$ too small can be inefficient, while $\tau$ too large will violate the linearization assumptions made in our discretization and cause flow instabilities.
We use a heuristic inspired by Armijo's condition \cite{nocedalwright} to choose the optimal step size,
\(\tau = \rho\clamp(\tau_*, \tau_\textrm{min}, \tau_\textrm{max})\),
where \(\rho = \frac{1}{n}\),
\begin{equation}
    \tau_* = -\frac{\rho(\A\V-\S)\cdot(\A\P) + 0.01 \lVert \P \rVert^2}{\rho \lVert \A\P \rVert^2}
    ,\quad
    \P = -\rho\A^\top(\A\V - \S)
    \;,
\end{equation}
and, by default, \(\tau_\textrm{min} = 10^{-6}\), \(\tau_\textrm{max} = 50\).

\subsection{Mesh resolution and quality} 
As mesh vertices $\mathbf{V}$ move during our flow, mesh quality rapidly degrades, producing degeneracies, flipped and thin triangles, and self-intersections. We solve this common problem of geometric flows by remeshing with the algorithm of \citet{botsch2004remeshing}, which uses a sequence of local improvement operations. After each flow iteration, we apply a single remeshing iteration using a given target edge-length $h$ (more iterations may be used if needed). We implement this operation in an output-sensitive manner, analogous to the approach of \citet{sellan2020opening}, by remeshing exclusively the regions of the surface that are the closest point on \(\mathbf{\Omega}\) to any of the \(\p_i\)
and violate the
SDF value \(s_i\) by more than a tolerance $\varepsilon$
(by default, \(5\cdot10^{-3}\) for 2D and \(10^{-2}\) for 3D).

Both our flow and our remeshing operations preserve the intrinsic shape's topology.
This restriction helps us avoid some of the most catastrophic failures of existing methods like Marching Cubes, which can produce wrongly disconnected mesh components at low resolutions (see Figs.~\ref{fig:teaser},~\ref{fig:armadillo}).
At the same time, it means that our starting surface mesh $\Omega^0$ needs to agree with the topology of the surface to be reconstructed.

Careful consideration must also be given to the mesh resolution during our flow, as encoded in the remesher's
target edge-length, \(h\).
As shown in \reffig{sharp-features}, our flow is capable of recovering much more faithful surface detail than existing grid-based methods.
Thus, we naturally wish to provide it with enough degrees of freedom (sufficiently low
target edge-length \(h\)) to accurately represent the surface.
At the same time, too many mesh vertices will make our flow iterations costly and the sphere reaching problem underconstrained. 

\begin{figure}
    \includegraphics{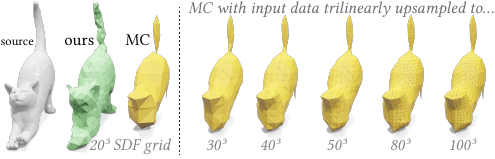}
\caption{
    Our algorithm's full exploitation of all the SDF input data means its improved performance against Marching Cubes is preserved even if one artificially upsamples the SDF input before using MC.
    }\label{fig:upsampled}
\end{figure}

\begin{wrapfigure}[12]{r}{0.25\columnwidth}
    \vspace{-15pt}
    \hspace{-19pt}
    \includegraphics{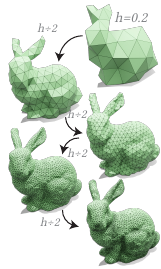}
\end{wrapfigure}
Ideally, then, we wish our flow to produce the lowest possible resolution mesh that can explain all SDF samples. We will achieve this by starting from a very high value of $h$ and running our flow until convergence, as identified by the energy's failure to decrease further than by a tolerance ($10^{-3}\varepsilon$) in the past 10 iterations, to obtain a coarse approximation of the surface (see inset). We will then halve $h$ and run our flow again, noting that our output-sensitive remesher will only refine the regions of the shape that are contributing to the energy (i.e., those that need the additional resolution). We repeat this process until a minimum $h_{min}$ is reached, by default set to be the average distance between SDF samples $\p_i$. Once $h_{\min}$ is reached, we run our flow until the energy has not decreased by more than $10^{-3}\varepsilon$ in the past \(100\) iterations.

\paragraph{Batching.} In practice, our algorithm's computational cost is dominated by the assembly of $\mathbf{A}$, which requires $n$ signed distance and closest point queries between each sphere origin and the current reconstruction mesh. Even though we manage to resolve each query in logarithmic time by assembling and using a bounding volume hierarchy, computing all $n$ queries can be costly. Thus, for large values of $n$ (e.g., $n>50^3$), we propose using randomly chosen batches of spheres at each iteration. Empirically, we note that interior spheres are more critical to our flow's stability; therefore, we only batch exterior spheres. By default, we make the batch size $\min(n,20000)$.

\begin{figure}
    \includegraphics{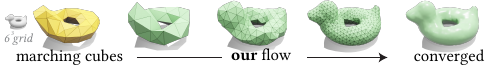}
\caption{
    Our flow is not limited to genus zero shapes, but the topology of the initial surface must match the reconstruction's. Marching Cubes may provide a good starting surface in these non-genus-zero cases.}\label{fig:non-genus-zero}
\end{figure}

\begin{figure}
\includegraphics{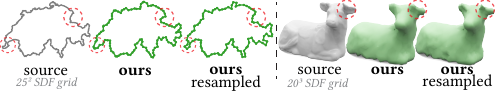}
\caption{We can further improve the reconstruction quality on coarse grids by adding just a few SDF samples after the flow has converged, due to the gridless nature of our method.
}\label{fig:resampling}
\end{figure}
  
\section{Results \& Experiments}

\paragraph{Implementation details.} We implemented our algorithm in Python using \textsc{Gpytoolbox} \cite{gpytoolbox} for common geometry processing subroutines including our flow's remesher as well as the Marching Cubes reconstruction \cite{lorensen1987marching} used in our comparisons. Our comparisons to Neural Dual Contouring \cite{chen2022neural} use the authors' publicly available implementation, including following their preprocessing instructions. We report timings on a 20-Core M1 Ultra Mac Studio with 128GB RAM. We rendered our figures in Blender, using \textsc{BlenderToolbox} \cite{blendertoolbox}.

Our method's complexity is determined by three algorithmic steps that must be executed at each flow iteration. First, the signed distances from each $\p_i$ to the current mesh are computed with a complexity of $\mathcal{O}((b+m)\log(m))$ (where $m$ is the number of current mesh vertices and $b$ is the batch size). Secondly, the linear system in Eq. (\ref{eq:discretetimestepping}) adds an $\mathcal{O}(m^{1+f})$ term, where $f$ accounts for the sparse system solve (we cannot take advantage of precomputation as the mesh changes in every iteration). Finally, our remeshing step adds an $\mathcal{O}(\tilde{m})$ term, where $\tilde{m}<m$ is the size of the active region of the current mesh. In the limit, this means each flow iteration will be asymptotically dominated by $\max(b\log(m),m^{1+f})$; in practice, we find this to be the case except for very low values of $m$ and $b$, where the constant factors in the remesher complexity dominate.

Our input shapes are scaled to fit the box $[-\frac12,\frac12]^n$, and our SDF grids are constructed in $[-1,1]^n$.
We use default parameters, unless otherwise specified in the supplemental material.
Unless otherwise specified, we initialize our examples with a unit icosahedral sphere, but our flow can handle other initializations (Fig.~\ref{fig:non-genus-zero}).

\begin{table*}
\begin{center}

    \resizebox{\linewidth}{!}{
    \begin{tabular}{c|ccc|ccc|ccc|c}
\rowcolor{LavenderBlue} Grid size & Hdf MC & Hdf NDCx  & Hdf Ours  & Chr MC  & Chr NDCx  & Chr Ours & $\ESDF$ MC & $\ESDF$ NDCx & $\ESDF$ Ours & Time Ours  \\[1pt]
\hline
$6^3$ & 0.3351 & 0.2597 & \textbf{0.1236} & 0.1918 & 0.1135 & \textbf{0.0569} & 31.4645 & 10.2969 & \textbf{0.1091} & 1.1214 \\[1pt]
\hline
\rowcolor[HTML]{EFEFEF}$10^3$ & 0.2518 & 0.1954 & \textbf{0.0846} & 0.1053 & 0.0662 & \textbf{0.0343} & 13.8933 & 6.2637 & \textbf{0.1152} & 1.2686 \\[1pt]
\hline
$20^3$ & 0.1486 & 0.1163 & \textbf{0.0631} & 0.0465 & 0.0311 & \textbf{0.0210} & 4.7667 & 2.8065 & \textbf{0.1377} & 4.7881 \\[1pt]
\hline
\rowcolor[HTML]{EFEFEF}$30^3$ & 0.0756 & 0.0494 & \textbf{0.0396} & 0.0206 & 0.0127 & \textbf{0.0118} & 0.6125 & 0.1451 & \textbf{0.0280} & 6.0939 \\[1pt]
\hline
$40^3$ & 0.0581 & \textbf{0.0366} & 0.0417 & 0.0143 & \textbf{0.0089} & 0.0100 & 0.3933 & 0.0910 & \textbf{0.0135} & 6.6929 \\[1pt]
\hline
\rowcolor[HTML]{EFEFEF}$50^3$ & 0.0501 & 0.0360 & \textbf{0.0253} & 0.0107 & 0.0077 & \textbf{0.0074} & 0.2451 & 0.1042 & \textbf{0.0050} & 9.4438     \end{tabular}}
    \end{center}
\caption{Across a diverse set of examples and resolutions, our flow exhibits lower Hausdorff ("Hdf"), Chamfer ("Chr") and SDF ($\ESDF$) errors than Marching Cubes, while surpassing or matching data-driven approaches like Neural Dual Contouring.
    Time given in seconds.
    Data averaged over
    Table \ref{table:dataset-comparison} (supplemental).
    }\label{table:average}
\end{table*}

\begin{figure}
\includegraphics{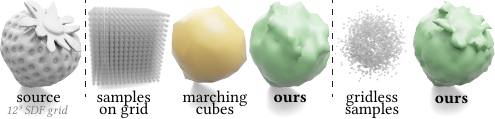}
\caption{SDF sampled from the same source
    on a
grid
    and with the same number of samples on a noisy point cloud of the source.
Alternate sampling strategies unavailable to grid-based methods allow us to recover more information with the same number of SDF samples.}\label{fig:gridless-samples}
\end{figure}

\subsection{Comparisons} The sole input to our algorithm is a set of query points $\p_i$ and corresponding SDF values $s_i$. In the specific case where these samples are placed on a structured grid, this input matches that of the timeless reconstruction algorithm Marching Cubes \cite{lorensen1987marching}. 
By allowing for the training on a vast dataset and the storing of a large number of network weights, recent advances like Neural Dual Contouring \cite{chen2022neural} have been shown to outperform most other reconstruction algorithms.

\paragraph{Qualitative comparisons.}
Throughout the paper, we qualitatively show our algorithm's improved performance against Marching Cubes (MC). At low resolutions, MC often produces little more than disconnected ``blobs'', often missing entire regions of the shape (see Figs.~\ref{fig:teaser},~\ref{fig:armadillo}). By contrast, our method shines at these resolutions, where exploiting all the global information provided by the SDF samples can recover features completely absent from the MC reconstruction (see Figs.~\ref{fig:sharp-features} and \ref{fig:gridless-samples}), an advantage that is preserved even if the data is upsampled artificially to denser grids (see \reffig{upsampled}). Even at higher resolutions, our global SDF-aware reconstruction captures significantly more detail (see Figs.~\ref{fig:high-resolution}, \ref{fig:hyperparameters} and~\ref{fig:large-grid}).

\begin{figure*}
    \includegraphics{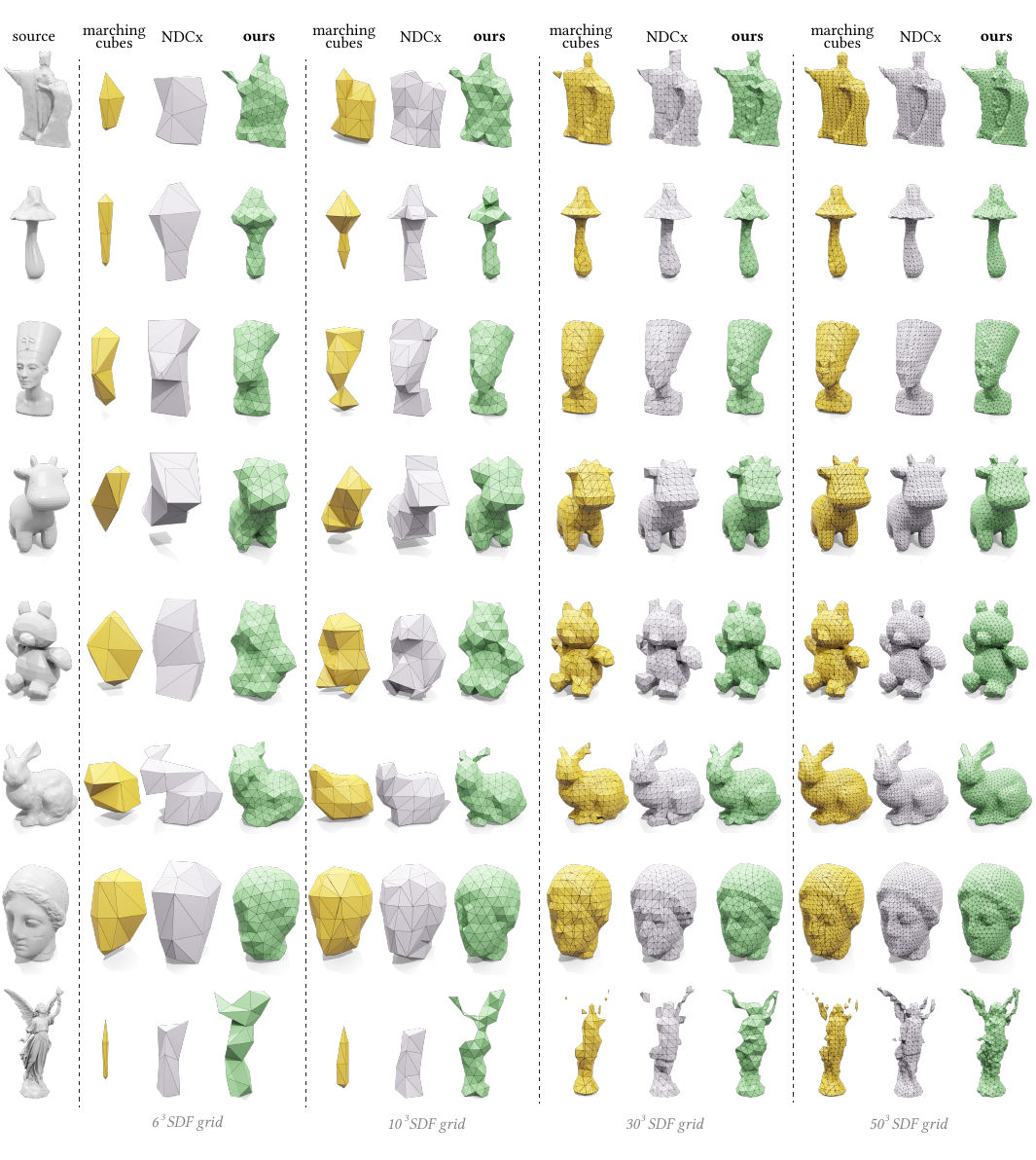}
    \caption{
    Our algorithm strikingly outperforms Marching Cubes and Neural Dual Contouring (NDCx) at low and medium resolutions.
    }\label{fig:big-comparison}
\end{figure*}

In Figs.~\ref{fig:teaser}, \ref{fig:armadillo}, and \ref{fig:big-comparison}, we additionally compare our algorithm's effectiveness with Neural Dual Contouring \cite{chen2022neural}. To make the comparison as generous as possible, we used the highest performing version of the authors' publicly available trained models~\cite{ndcgit},
NDCx, which combines their network with elements of their previous work's learned model \cite{chen2021neural}. Even though their data-driven approach manages to outperform Marching Cubes in almost all our tests, we qualitatively find that our purely geometric algorithm consistently outperforms both at low and medium resolutions despite requiring no training.

\paragraph{Quantitative comparisons.}
Inspired by the evaluations in the work by \citet{chen2022neural}, we also compare our algorithm's performance quantitatively. In Fig.~\ref{fig:big-comparison}, we run our algorithm using its default parameters as well as Marching Cubes and NDCx on shapes from a diverse set of origins whose SDFs have been sampled at different resolutions.
In our supplemental material, we attach a table comparing the Hausdorff distance to the ground truth mesh as well as Chamfer distance and our own SDF energy $\ESDF$, while Table~\ref{table:average} shows the average values for each resolution.
While placing fewer requirements on the input (any set of points $\p_i$ versus a structured grid), our algorithm consistently outperforms Marching Cubes across the board, often by several integer factors.
While requiring no training, our algorithm surpasses NDCx at low resolutions and remains competitive at medium and higher resolutions.
We note that MC requires between $20^3$ and $30^3$ SDF grid samples to match the accuracy of our algorithm at the lowest of resolutions ($6^3$ grid). Thus, our algorithm reduces memory storage requirements for equal surface accuracy by a factor of between $37$ and $125$.

\begin{figure}
\includegraphics{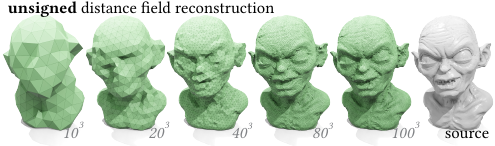}
\caption{
    By relaxing the constraints in our method, our flow can be seamlessly applied to \emph{unsigned} distance fields at diverse resolutions.}\label{fig:unsigned}
\end{figure}

\subsection{Parameters}
A number of parameters affect our method's ability to extract all information from its SDF input.
Among these, the most crucial is the minimum mesh edge-length $h_{\min}$.
Choosing $h_{\min}$ too high can cause the method to miss information available in the SDF samples.
A very small $h_{\min}$ can negatively affect performance, and also underconstrain the problem, leading to (completely valid) solutions that have high-frequency noise.
Our remeshing procedure contains a regularization step that combats this noise, but does not completely obviate it.
Empirically, we find that setting $h_{\min}$ to be the average closest-distance between samples $\p_i$ (i.e., the gridless analogue of the grid edge-length) is a useful heuristic, whose reliability we show across resolutions in \reffig{big-comparison}, Table~\ref{table:average} and the supplemental.

\subsection{SDF sampling}
While many algorithms rely on SDF samples to be located on a structured (regular or not) grid, our method is completely agnostic to the position of the samples $\p_i$.
We can take advantage of this in multiple ways. For example, we can run our algorithm, unchanged, on SDF data sampled on fully unstructured point clouds, exploiting prior information (Fig.~\ref{fig:gridless-samples}). In settings where the source SDF \emph{function} is available to be queried, we can add more samples \((\p_i,s_i)\) after our method has converged, and run it from the previous result (Fig.~\ref{fig:resampling}) to incrementally improve the reconstruction.
Our heuristic for adding samples is to generate \(m_\textrm{trial}\) samples on \(\mathbf{\Omega}\),
randomly displace them in the normal direction by a normal distribution scaled by \(0.05\),
and select the \(m_\textrm{new}\) samples farthest away from the surface of any SDF sphere (but at most one per mesh element).
By default, \(m_\textrm{new}=2\sqrt{n}\) in 2D, \(m_\textrm{new}=2\sqrt[3]{n}\) in 3D, and \(m_\textrm{trial} = 50 m_\textrm{new}\).

\subsection{Beyond SDFs} 

Signed Distance Fields are a powerful representation that we have shown can be exploited to obtain a surprisingly large amount of information about a given object. Often, however, computing exact SDFs can be costly or impracticable, forcing one to relax some of the assumptions in the traditional SDF definition.

Consider the case of \emph{unsigned} distance fields, which lack the inside-outside information contained in the sign of traditional SDFs. As shown in \reffig{unsigned}, our flow can very easily be employed to reconstruct meshes from these functions, merely by always making $\sigma_i=1$  in \refequ{ti}. Intuitively, this means we move the surface towards the closer of the sphere's two possible tangent points, $\t_i$.

\begin{figure}
\includegraphics{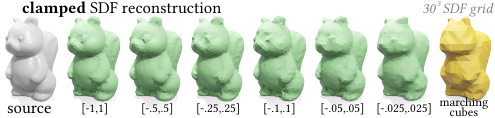}
\caption{
    \emph{Clamped} or \emph{truncated} SDFs discard information our method needs to capture the shape's detail, but our method degrades gracefully and still outperforms Marching Cubes even for aggressive clamping parameters.}\label{fig:clamped}
\end{figure}
\begin{figure}
\includegraphics{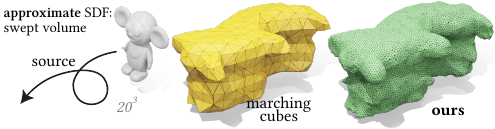}
\caption{
    A \emph{swept volume} SDF is accurate only outside the object. Inside, it is only a bound on distance. By relaxing its assumptions, we can use our method for swept volume reconstruction.}\label{fig:sv}
\end{figure}

Another relaxation of SDFs are \emph{clamped}, \emph{truncated}, or \emph{narrow band} SDFs, that take a constant value at spatial positions ``sufficiently far'' from the surface. This is a common representation in highly performant modelling applications and, more recently, in machine learning models when one wants to focus learning near the object's surface (see, e.g., \cite{park2019deepsdf}). Generalizing our algorithm to these representations is conceptually simple: for a given clamp value $\sigma_c$, we allow the tangency requirement (but not the intersection-free one) to be violated for those spheres with radii
larger than
$\sigma_c$. In practice, this amounts to zeroing out the $i$-th row of $\A$ if $|\phi(\p_i,\Omega^t)|>|s_i|>\sigma_c$. Our algorithm relies on faraway spheres to provide additional information about the reconstruction; therefore, using a clamped SDF necessarily results in a progressive loss of detail (see \reffig{clamped}). 

Yet another common SDF-based representation is formed by instead providing bounds on the true shape's signed distance (these are referred to as \emph{conservative} SDFs by \citet{takikawa2022dataset}). Such SDFs appear naturally as the output of Boolean operations on signed distance functions. A recently studied example of this are \emph{swept volumes}, which can be represented by taking the minimum of the SDF of an object along a trajectory; however, this representation is only an exact SDF \emph{outside} the volume, while only a bound inside. As we show in \reffig{sv}, all that is needed to apply our flow to swept volume reconstruction is to relax the tangency constraint of the negative-sign spheres only. In practice, this amounts to zeroing the $i$-th row of $A$ if $|\phi(\p_i,\Omega^t)|>|s_i|$ and $s_i<0$. We believe this to be a promising application of our work, as swept volume approximate SDFs are often extremely costly to query \cite{sellan2021swept}.

\section{Discussion and Conclusions}
We have leveraged our new tangent-spheres interpretation to develop an effective isosurfacing method, called \emph{Reach for the Spheres}, that exploits the full representational power of SDFs.
Using only geometric information present in a standard discrete SDF, we are able to recover noticeably more detail than previous general-purpose isosurfacing schemes.
Our method especially shines on low-resolution SDF grids, where it is able to exploit every last bit of information that other methods might miss, and it can match traditional methods for high resolutions. By releasing our method to the Graphics community, we hope to renew interest in lightweight, low-resolution SDF representations and enable novel, scalable applications.

\begin{figure}
\includegraphics{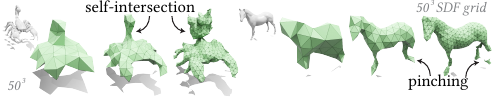}
\caption{Like many geometric flows, our algorithm can occasionally produce singularities, corresponding to attempts to dynamically change topology.}\label{fig:singularity}
\end{figure}

\begin{figure}
\includegraphics{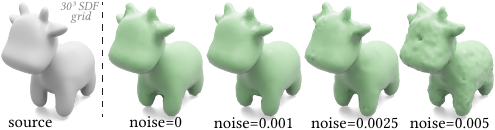}
\caption{
        Adding Gaussian noise to the SDF input values, with increasing standard deviation.
    For small values, our flow degenerates gracefully.
    At a standard deviation of 0.005 (i.e., 0.5\% of the shape's bounding box length), our flow hits a singularity before the stopping criterion is reached.
    }\label{fig:noise}
\end{figure}

Our method is not yet robust to self-intersections, nor does it support topology changes, as needed to straightforwardly handle difficult multi-component or nonzero genus shapes.
Thus, as a limitation, our method can exhibit self-intersection and pinching effects due to singularities in the discrete flow (\reffig{singularity}). 
Our algorithm's current inability to handle these singularities also limits its efficacy on noisy SDF data (see \reffig{noise}).
We are optimistic that existing mesh-based fluid simulation surface tracking techniques \cite{wojtan2011liquid} can help overcome these restrictions.

Furthermore, a surface that perfectly satisfies a given discrete SDF can often still have significant flexibility at the finer scales; an exciting direction is to incorporate specific priors for particular applications, via additional regularization or data-driven approaches.
There is also ample room to improve the performance of our method, by using more elaborate methods for closest point computations, solving linear equations, and remeshing.

Beyond surface reconstruction, a vast array of other graphics techniques rely on discrete SDFs across simulation, geometry processing, and rendering.
We look forward to exploring whether incorporating the tangent-spheres perspective can yield comparable improvements for these applications as well.

 \section*{Acknowledgements}

This work is funded in part by the Natural Sciences and Engineering Research Council of Canada (Grant RGPIN-2021-02524), an NSERC Vanier Scholarship and an Adobe Research Fellowship.

We thank Abhishek Madan for technical help and for proofreading.
We thank Aravind Ramakrishnan, and Hsueh-Ti Derek Liu for proofreading.
The first author would also like to thank the second and third authors for inviting her to visit their respective institutions, which served as the foundation for this collaboration.

We acknowledge the authors of the 3D models used throughout this paper and thank them for making them available for academic use.
Figures in this work contain the
koala \cite{koala-mesh},
springer \cite{springer-mesh},
tower \cite{tower-mesh},
bunny \cite{bunny-mesh},
turtle \cite{turtle-mesh},
armadillo \cite{armadillo-mesh},
cat \cite{cat-mesh},
duck \cite{duck-mesh},
cow \cite{cow-mesh},
strawberry \cite{strawberry-mesh},
plush toy \cite{plush-toy-mesh},
spot \cite{spot-mesh},
scorpion \cite{scorpion-mesh},
mushroom \cite{mushroom-mesh},
Nefertiti \cite{nefertiti-mesh},
Max Planck \cite{max-mesh},
Igea \cite{igea-mesh},
horse \cite{horse-mesh},
Argonath \cite{argonath-mesh},
squirrel \cite{squirrel-mesh},
teddy \cite{teddy-mesh}
and
Lucy \cite{lucy-mesh}
meshes.

\bibliographystyle{ACM-Reference-Format}
\bibliography{references.bib}

\clearpage
\appendix

\setcounter{page}{1}
\section*{Supplemental Material}

\subsection*{Parameters}

In this section we list \(h_\textrm{min}\) and non-default parameters used in this article.

\paragraph{Fig.~\ref{fig:teaser}.}
\(h_\textrm{min} = 0.02\) for grid size \(10\).
\(h_\textrm{min} = 0.01\) for grid size \(50\).
\(\varepsilon = 10^{-4}\).
\(3\) remesh iterations.

\paragraph{Fig.~\ref{fig:sharp-features}.}
\(h_\mathrm{min}=0.03\).
\(t_\mathrm{max}=10\) in 3D.

\paragraph{Fig.~\ref{fig:flow-2d}.}
\(h_{\mathrm{min}}=0.01\).

\paragraph{Fig.~\ref{fig:flow-3d}.}
\(h_{\mathrm{min}}=0.02\).
Batching turned off.

\paragraph{Fig.~\ref{fig:high-resolution}.}
\(h_\textrm{min}=0.008\).

\paragraph{Fig.~\ref{fig:armadillo}.}
Left to right: \(h_{\min}=0.05\), \(0.02\) and \(0.008\).

\paragraph{Fig.~\ref{fig:large-grid}.}
For grid size \(n^3\),
\(h_\mathrm{min}=\frac{2}{n}\), and
\(\varepsilon=\frac{10^{-2}}{n}\).

\paragraph{Fig.~\ref{fig:resampling}.}
\(h_\mathrm{min}=0.035\) in 2D; \(h_\mathrm{min}=0.04\) in 3D.

\paragraph{Fig.~\ref{fig:gridless-samples}.}
\(h_\mathrm{min}=0.06\).
\(\varepsilon=10^{-3}\)

\paragraph{Fig.~\ref{fig:noise}.}
\(h_\mathrm{min}=0.015\).
\vfill\null

\vspace{600pt}

\subsection*{Detailed Quantitive Evaluation Data}
Table \ref{table:dataset-comparison} contains the detailed results of our quantitative evaluations for the SDF reconstruction problem on a variety of shapes, comparing our method with Marching Cubes and NDCx.

\begin{table*}[t]
    \begin{center}
    \small{
\resizebox{\linewidth}{!}{
    \begin{tabular}{l|c|ccc|ccc|ccc|c}
      Shape & $n$ & Hdf MC & Hdf NDCx  & Hdf Ours  & Chr MC  & Chr NDCx  & Chr Ours & $\ESDF$ MC & $\ESDF$ NDCx & $\ESDF$ Ours & Time Ours  \\
\hline
mushroom & 6 & 0.1952 & 0.1489 & \textbf{0.1417} & 0.1274 & 0.1103 & \textbf{0.0753} & 11.7905 & 1.1676 & \textbf{0.0519}&0.5147 \\
\hline
mushroom & 10 & 0.1221 & \textbf{0.0819} & 0.1023 & 0.0715 & 0.0456 & \textbf{0.0420} & 3.9187 & 0.4795 & \textbf{0.0455}&1.1983 \\
\hline
mushroom & 20 & 0.0611 & \textbf{0.0485} & 0.0606 & 0.0345 & \textbf{0.0197} & 0.0200 & 0.5752 & 0.1567 & \textbf{0.0519}&6.0584 \\
\hline
mushroom & 30 & 0.0546 & \textbf{0.0233} & 0.0407 & 0.0215 & \textbf{0.0114} & 0.0116 & 0.3133 & 0.0758 & \textbf{0.0150}&3.9053 \\
\hline
mushroom & 40 & 0.0532 & \textbf{0.0217} & 0.0238 & 0.0165 & 0.0077 & \textbf{0.0077} & 0.2285 & 0.0286 & \textbf{0.0092}&4.1968 \\
\hline
mushroom & 50 & 0.0382 & \textbf{0.0186} & 0.0188 & 0.0120 & 0.0065 & \textbf{0.0062} & 0.1012 & 0.0128 & \textbf{0.0038}&7.1066 \\
\hline
nefertiti & 6 & 0.2007 & 0.1296 & \textbf{0.0988} & 0.1384 & 0.0671 & \textbf{0.0462} & 11.9997 & 0.6109 & \textbf{0.0520}&1.8593 \\
\hline
nefertiti & 10 & 0.1826 & 0.0760 & \textbf{0.0645} & 0.0976 & 0.0379 & \textbf{0.0188} & 5.1541 & 0.3250 & \textbf{0.0427}&1.6463 \\
\hline
nefertiti & 20 & 0.0744 & 0.0419 & \textbf{0.0381} & 0.0304 & 0.0146 & \textbf{0.0142} & 0.9196 & 0.0742 & \textbf{0.0416}&6.6072 \\
\hline
nefertiti & 30 & 0.0406 & 0.0382 & \textbf{0.0286} & 0.0144 & \textbf{0.0080} & 0.0086 & 0.2235 & 0.0414 & \textbf{0.0145}&10.2500 \\
\hline
nefertiti & 40 & 0.0382 & 0.0287 & \textbf{0.0246} & 0.0119 & \textbf{0.0063} & 0.0069 & 0.1655 & 0.0190 & \textbf{0.0100}&6.4140 \\
\hline
nefertiti & 50 & 0.0341 & 0.0316 & \textbf{0.0264} & 0.0076 & \textbf{0.0056} & 0.0059 & 0.0459 & 0.0128 & \textbf{0.0032}&11.1034 \\
\hline
bunny & 6 & 0.2947 & \textbf{0.1669} & 0.1769 & 0.1997 & 0.0938 & \textbf{0.0626} & 15.2121 & 1.9051 & \textbf{0.0438}&0.9502 \\
\hline
bunny & 10 & 0.4295 & 0.4051 & \textbf{0.0860} & 0.1199 & 0.0982 & \textbf{0.0332} & 16.6344 & 11.3773 & \textbf{0.0685}&1.3017 \\
\hline
bunny & 20 & 0.1975 & 0.1348 & \textbf{0.0612} & 0.0453 & 0.0271 & \textbf{0.0182} & 3.6855 & 1.0939 & \textbf{0.0527}&4.2061 \\
\hline
bunny & 30 & 0.0580 & \textbf{0.0383} & 0.0397 & 0.0158 & 0.0110 & \textbf{0.0097} & 0.2503 & 0.0349 & \textbf{0.0226}&8.9931 \\
\hline
bunny & 40 & 0.0638 & \textbf{0.0273} & 0.0351 & 0.0131 & 0.0086 & \textbf{0.0080} & 0.1889 & 0.0196 & \textbf{0.0097}&10.1580 \\
\hline
bunny & 50 & 0.0487 & 0.0309 & \textbf{0.0219} & 0.0090 & 0.0069 & \textbf{0.0063} & 0.0789 & 0.0096 & \textbf{0.0035}&14.7060 \\
\hline
argonath & 6 & 0.3763 & 0.2707 & \textbf{0.1146} & 0.2341 & 0.0827 & \textbf{0.0393} & 42.1481 & 6.2328 & \textbf{0.1402}&0.7281 \\
\hline
argonath & 10 & 0.3271 & 0.2104 & \textbf{0.0751} & 0.0919 & 0.0548 & \textbf{0.0314} & 12.6527 & 5.5259 & \textbf{0.2200}&0.9053 \\
\hline
argonath & 20 & 0.1707 & 0.1538 & \textbf{0.0486} & 0.0482 & 0.0334 & \textbf{0.0214} & 4.4039 & 1.7451 & \textbf{0.0794}&5.2204 \\
\hline
argonath & 30 & 0.0953 & 0.0401 & \textbf{0.0320} & 0.0197 & 0.0120 & \textbf{0.0119} & 0.8446 & 0.1481 & \textbf{0.0318}&4.0973 \\
\hline
argonath & 40 & 0.0513 & \textbf{0.0282} & 0.0303 & 0.0149 & \textbf{0.0084} & 0.0089 & 0.3563 & 0.0245 & \textbf{0.0139}&6.2986 \\
\hline
argonath & 50 & 0.0443 & \textbf{0.0237} & 0.0274 & 0.0097 & \textbf{0.0065} & 0.0067 & 0.2083 & 0.0196 & \textbf{0.0067}&8.0372 \\
\hline
lucy & 6 & 0.4025 & 0.3718 & \textbf{0.0974} & 0.1601 & 0.1128 & \textbf{0.0549} & 45.0055 & 19.7607 & \textbf{0.5921}&0.3424 \\
\hline
lucy & 10 & 0.4490 & 0.4405 & \textbf{0.1046} & 0.1599 & 0.1356 & \textbf{0.0499} & 44.8871 & 29.1435 & \textbf{0.4019}&0.7341 \\
\hline
lucy & 20 & 0.4054 & 0.3611 & \textbf{0.0937} & 0.1161 & 0.0943 & \textbf{0.0349} & 27.0360 & 20.5396 & \textbf{0.8438}&4.6527 \\
\hline
lucy & 30 & 0.1243 & 0.1190 & \textbf{0.0757} & 0.0358 & 0.0251 & \textbf{0.0232} & 1.5015 & 0.4685 & \textbf{0.0952}&4.6289 \\
\hline
lucy & 40 & 0.1162 & 0.1166 & \textbf{0.0534} & 0.0258 & 0.0183 & \textbf{0.0164} & 1.3426 & 0.4506 & \textbf{0.0531}&6.1530 \\
\hline
lucy & 50 & 0.1409 & 0.1176 & \textbf{0.0380} & 0.0227 & 0.0178 & \textbf{0.0140} & 1.4041 & 0.8705 & \textbf{0.0117}&5.4008 \\
\hline
igea & 6 & 0.1716 & 0.1026 & \textbf{0.0602} & 0.1192 & 0.0561 & \textbf{0.0218} & 4.3137 & 1.1548 & \textbf{0.0158}&1.2822 \\
\hline
igea & 10 & 0.1078 & 0.0727 & \textbf{0.0470} & 0.0554 & 0.0294 & \textbf{0.0164} & 1.3410 & 0.3004 & \textbf{0.0119}&2.3998 \\
\hline
igea & 20 & 0.0653 & \textbf{0.0421} & 0.0569 & 0.0203 & 0.0149 & \textbf{0.0143} & 0.2639 & 0.0748 & \textbf{0.0151}&4.3184 \\
\hline
igea & 30 & 0.0386 & \textbf{0.0276} & 0.0292 & 0.0116 & 0.0093 & \textbf{0.0087} & 0.0687 & 0.0237 & \textbf{0.0047}&6.2384 \\
\hline
igea & 40 & 0.0296 & \textbf{0.0184} & 0.0237 & 0.0089 & \textbf{0.0071} & 0.0074 & 0.0368 & 0.0061 & \textbf{0.0023}&7.4121 \\
\hline
igea & 50 & 0.0259 & \textbf{0.0136} & 0.0207 & 0.0075 & \textbf{0.0065} & 0.0067 & 0.0222 & 0.0049 & \textbf{0.0017}&9.0712 \\
\hline
armadillo & 6 & 0.5501 & 0.4145 & \textbf{0.1573} & 0.2729 & 0.1810 & \textbf{0.0911} & 78.0223 & 33.4996 & \textbf{0.0651}&1.0500 \\
\hline
armadillo & 10 & 0.3669 & 0.2590 & \textbf{0.1167} & 0.1816 & 0.1028 & \textbf{0.0490} & 35.4634 & 11.6023 & \textbf{0.1679}&1.0748 \\
\hline
armadillo & 20 & 0.1925 & 0.1766 & \textbf{0.0495} & 0.0658 & 0.0488 & \textbf{0.0209} & 6.9130 & 3.7000 & \textbf{0.0893}&4.1866 \\
\hline
armadillo & 30 & 0.1382 & 0.0931 & \textbf{0.0396} & 0.0283 & 0.0189 & \textbf{0.0129} & 1.4777 & 0.4933 & \textbf{0.0418}&5.0136 \\
\hline
armadillo & 40 & 0.1067 & \textbf{0.0624} & 0.1335 & 0.0191 & \textbf{0.0116} & 0.0221 & 1.2729 & 0.2819 & \textbf{0.0149}&2.7483 \\
\hline
armadillo & 50 & 0.0770 & 0.0573 & \textbf{0.0302} & 0.0125 & 0.0091 & \textbf{0.0085} & 0.3559 & 0.0883 & \textbf{0.0083}&8.8009 \\
\hline
teddy & 6 & 0.2284 & 0.1726 & \textbf{0.1407} & 0.1369 & 0.0836 & \textbf{0.0587} & 11.9344 & 2.0409 & \textbf{0.0347}&1.7151 \\
\hline
teddy & 10 & 0.1692 & 0.1615 & \textbf{0.1039} & 0.0922 & 0.0655 & \textbf{0.0418} & 4.4468 & 1.6534 & \textbf{0.0409}&1.2807 \\
\hline
teddy & 20 & 0.1280 & 0.1122 & \textbf{0.0821} & 0.0363 & 0.0262 & \textbf{0.0255} & 1.0904 & 0.4025 & \textbf{0.0473}&4.4172 \\
\hline
teddy & 30 & 0.0480 & \textbf{0.0279} & 0.0353 & 0.0177 & 0.0115 & \textbf{0.0113} & 0.1661 & 0.0561 & \textbf{0.0144}&7.2587 \\
\hline
teddy & 40 & 0.0410 & \textbf{0.0232} & 0.0473 & 0.0117 & \textbf{0.0083} & 0.0090 & 0.0641 & 0.0361 & \textbf{0.0067}&5.6651 \\
\hline
teddy & 50 & 0.0343 & \textbf{0.0266} & 0.0275 & 0.0092 & \textbf{0.0068} & 0.0070 & 0.0284 & 0.0093 & \textbf{0.0031}&8.0650 \\
\hline
spot & 6 & 0.2901 & 0.3463 & \textbf{0.1325} & 0.2069 & 0.1743 & \textbf{0.0619} & 24.3347 & 11.6865 & \textbf{0.0313}&1.2620 \\
\hline
spot & 10 & 0.2188 & 0.1574 & \textbf{0.0654} & 0.1177 & 0.0621 & \textbf{0.0291} & 10.7422 & 2.1484 & \textbf{0.0498}&1.0666 \\
\hline
spot & 20 & 0.0932 & \textbf{0.0451} & 0.0508 & 0.0313 & \textbf{0.0176} & 0.0189 & 1.0994 & 0.2004 & \textbf{0.0602}&4.6396 \\
\hline
spot & 30 & 0.0654 & 0.0474 & \textbf{0.0465} & 0.0158 & 0.0101 & \textbf{0.0092} & 0.2841 & 0.0871 & \textbf{0.0135}&4.3807 \\
\hline
spot & 40 & 0.0320 & \textbf{0.0209} & 0.0231 & 0.0095 & 0.0069 & \textbf{0.0063} & 0.0520 & 0.0293 & \textbf{0.0043}&7.5936 \\
\hline
spot & 50 & 0.0251 & 0.0264 & \textbf{0.0180} & 0.0077 & 0.0059 & \textbf{0.0056} & 0.0302 & 0.0112 & \textbf{0.0021}&11.0150

     \end{tabular}}
    \caption{Quantitative Evaluation Data
    \label{table:dataset-comparison}}
    }
    \end{center}
\end{table*}

\end{document}